\documentclass[12pt,floats]{article}
\usepackage{epsfig}
\begin{document}

\title{\bf BEC for photons and neutral pions}

\author{Oleg V. Utyuzh and Grzegorz Wilk
\\[2ex]  
{\it Nuclear Theory Department,} \\
{\it The Andrzej So\l tan Institute for Nuclear Studies,}\\
{\it Ho\.za 69, 00681 Warsaw, Poland,} \\
{\it e-mail: utyuzh@fuw.edu.pl and wilk@fuw.edu.pl}}

\date{\today}
\maketitle

\begin{abstract}
The role of BEC for photons and neutral pions is briefly discussed
and illustrated by means of new method of numerical modelling of BEC
proposed by us recently.\\

\end{abstract}

\newpage

Photons and $\pi^0$'s play particular role in high energy
multiparticle processes, especially in heavy ion collisions, which
are believed to be the place where the Quark Gluon Plasma is possibly
formed. This is because they are immune to the Coulomb final state
interactions distorting BEC analysis of charged particles. The
$\pi^0$ interferometry can be then used as a kind of check allowing
to estimate the correcteness of procedures accounting for this effect
in BEC between charged particles. Of minor experimental consequensce
at present (but otherwise very interesting) is fact that, because
both photons and $\pi^0$'s are also their own antiparticles, one
expects that for low momenta (actually, for $k_{\gamma, \pi^0}
\rightarrow 0$) one should observe some additional effects emerging
from the particle-antiparticle correlation, a pure quantum mechanical
effect discussed at length in \cite{Weiner}. Photons are also the
only particles which could, in principle, provide us with detailed
picture of the whole history of heavy ion collision process because
they are produced anywhere where one has deceleration or acceleration
of charges whereas hadrons (including $\pi^0$'s) are produced only at
final hadronization period of collision. This means that they are
produced starting with very early stage of collision (sometimes called
"preequilibrium"), through the stage of formation and equilibration
of the QGP and through the mixed stage where first hadrons are being
produced in QGP environment and ending with the last kinetic
freez-out stage where hadrons are finally formed (see \cite{Photons}
for most recent review; no BEC discussion is presented there,
however). It is obvious that production mechanism of photons at
different periods of collision will be different, ranging from
bremstrahlung at the beginning, via thermal emission at QGP phase, to
decays of neutral particles (mostly $\pi^0$'s) at the end
\cite{Photons}. For us the most interesting would be directly
produced photons, especially thermal ones originated from the QGP. 
However, according to the recent WA98 data \cite{Data}, they consist
of only a few percent of the all photons visible.

Photons should be also copiously produced at PHOS detector of ALICE
at LHC, therefore it is fully justified to prepare for this
possibility, especially for the possible BEC measurements, both with
direct and decay photons (for references see \cite{GW}, the most
recent paper on this subject is \cite{P} where updated references can
be found). Whereas the method  to be used for BEC are essentially the
same as for other particles (cf. \cite{GW,Weiner}) there are some new
problems one is facing which are connected with the fact that at
ALICE one probably will not be able to measure directly $\pi^0$'s but
only photons they are decaying to. Question which arises then is:
notwithstanding this fact, can one deduce BEC characteristics of
$\pi^0$'s from BEC characteristics of $\gamma's$? This problem has
been addressed in \cite{DS} and positive conclusion has been reached
as well as the procedure of such analysis has been provided, which
can easily be applied when data will be available. However, opposite
conclusion has been reached in \cite{P} where photons from $\pi^0$'s
were demonstrated to show very characteristic structure at small
values of $Q$ (of the order of pion mass) and flat distribution for
larger values of $Q$. On the other hand, it was also shown there
that, at least for SPS and RHIC energies, one can actually separate
correlations of photons produced directly in the hot zone and
residual correlations of decay photons (i.e., those coming
predominantly from $\pi^0$'s). It turns out that (at least in the
framework of the hydrodynamical model of collision used there) the
former contribute essentially to the region of small invariant
relative momenta, $Q_{inv} \leq 50$ MeV, while the later dominate the
large relative momomentu.  

Here we shall present numerical calculations in which $\pi^0$'s were
created by Monte Carlo algorithm, which incorporates already BEC
effect \cite{UW}. Referring to \cite{UW} for details and physical
motivation we shall concentrate here on the same problem as was
considered analytically in \cite{P} and \cite{DS}, namely, to what
extent BEC picture of original $\pi^0$ is visible in finally observed
$\gamma$'s to which they decay and what is the chance to detect BEC
of originally produced $\gamma$'s as well. To this end we have to
modify slightly algorithm used in \cite{UW} by allowing for varying
transverse momenta of produced $\pi^0$'s (they were fixed as
$p_T=\langle p_T\rangle$ in \cite{UW}, i.e., the case considered was
essentially a purely one-dimensional). It will be done in the
following way. After choosing energy $E$ of selected particles (done
in the same way as before \cite{UW}) we shall notice that $E =
\sqrt{p_T^2 + p_L^2 + \mu^2}$ (where $\mu$ denotes pion mass) and
choose now $p_T= |\vec{p}_T|$ from distribution $P(p_T) = \exp( -
p_T/\langle p_T\rangle)$ (in calculations presented here parameter
$\langle p_T\rangle = 0.2$ GeV/c) such that $p_T\in (0., \sqrt{E^2 -
\mu^2})$. Then the azimuthal angle $\phi \in [0, 2\pi]$ was selected
from uniform distribution and direction "right-left" for longitudinal
momentum $p_L$ was chosen. In this way $p_L=\pm
\sqrt{E^2-p_T^2-\mu^2}$. In this way we star a new elementary
emitting cell (cf. \cite{UW} for more details) and add to it other
particles of the same kind momenta of which were selected from
gaussian distributions with widths $\sigma_L$ and $\sigma_T$ treated
as free parameters. It turnes out that in the example considered here
results are not particularly sensitive to their difference therefore
we have put them equal: $\sigma_L=\sigma_T$ \footnote{The reason is
that our example is still very much one-dimensional. The reason is
that in \cite{UW} we wanted to check whether it is possible {\it in
principle} to reproduce $e+e^-$ annihilation data and their phase
space is clearly a quasi-one-dimensional. However, this limitation
does not affect results and conclusions presented here.}.

\begin{center}
\begin{tabular}{|cc|}
\hline
\begin{minipage}{7cm}
\begin{center}
\includegraphics[height=7cm,width=7cm]{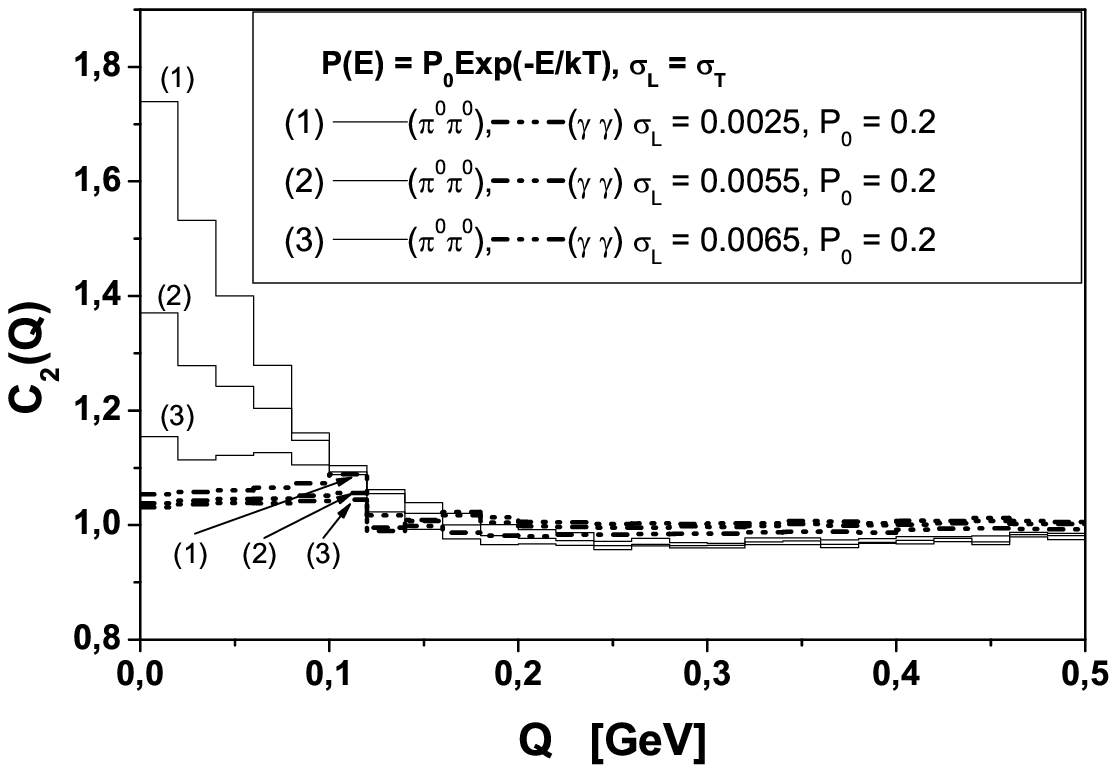} 
\end{center}
\end{minipage} &
\begin{minipage}{7cm}
\begin{center}
\includegraphics[height=7cm,width=7cm]{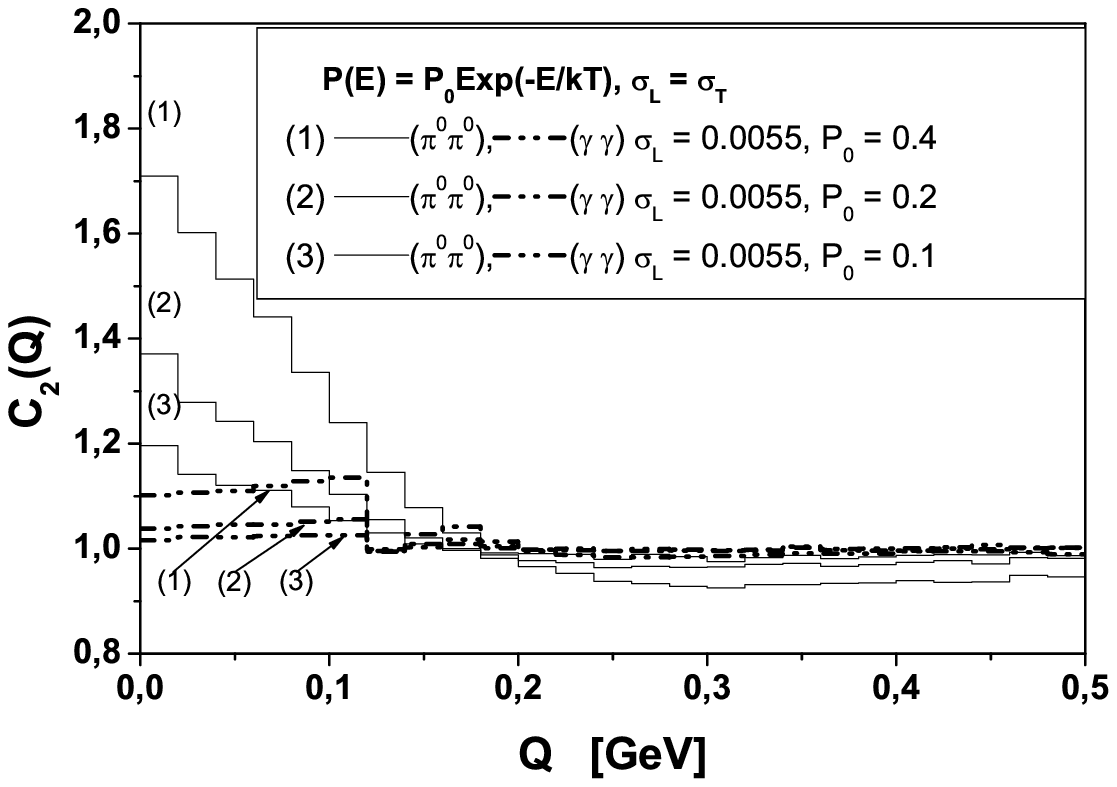} 
\end{center}
\end{minipage} \\
\hline
\end{tabular} \\ 
\vspace{0.5cm}{\scriptsize Fig.1 Example of BEC results for $\pi^0$
as produced by our algorithm (full lines) and BEC for $\gamma$'s
resulting from their decays. Left panel shows results for different
sizes of elementary emitting cells (EEC's) characterized by parameter
$\sigma_{L,T}$. Right panel shows results for different strength of
BEC effect characterized by parameter $P_0$ (leading to different
occupancy of the average EEC, cf. \cite{UW}).}
\end{center}

Our results are shown in Fig. 1. They confirm results of \cite{P},
namely that photons from decaying pions do 
not follow their BEC pattern but show, instead, their own pattern,
which is quite robust against different choices of parameters, and
which consist in essentially constant (but small) correlations for
small relative momenta vanishing after $Q$ of the order of pion mass.
The possible explanation of this result seems to be the same as in
\cite{P} (modulo possible deviations in numbers caused by our
limitation in $p_T$ and essentially $3$-dimensional case considered
in \cite{P})\footnote{At this moment we cannot offer any explanation
of difference between results of \cite{P} and \cite{DS} noticing only
that we do not think that explanation provided in \cite{P} and
attributing this difference to different way of introducing cuts in
momenta should not work for the first panel of figure presented in
\cite{DS}, where there is no limitation and result is still different
from that presented in \cite{P}.}.

\end{document}